\documentstyle[graphicx,prl,aps]{revtex}
   \newcommand{\vecbm}[1]{\mbox{$\boldmath#1$}}

   \newcommand{\lora} {{\boldmath$\longrightarrow$}}
\begin{document}
\paperwidth =15cm
%%\twocolumn
% \draft command makes pacs numbers print
\draft
%\tighten
\title{Second Law of Thermodynamics, Macroscopic Observables\\
within Boltzmann's Principle but without Thermodynamic Limit} 
% repeat the \author\address pair as needed
\author{D.H.E. Gross} \address{ Hahn-Meitner-Institut
  Berlin, Bereich Theoretische Physik,Glienickerstr.100\\ 14109
  Berlin, Germany and Freie Universit{\"a}t Berlin, Fachbereich
  Physik; \today} 
\maketitle
\begin{abstract}
  Boltzmann's principle $S=k\ln W$ allows to extend equilibrium
  thermo-statistics to ``Small'' systems {\em without invoking the
    thermodynamic limit} \cite{gross174,gross173,gross175}. Here we
  show that the formulation of non-equilibrium statistics and the
  Second Law can be easier formulated than by conventional theory. A
  deeper and more transparent understanding is thus possible. The main
  clue is to base statistical probability on {\em ensemble averaging}
  and {\em not} on time averaging. It is argued that due to the
  incomplete information obtained by macroscopic measurements
  thermodynamics handles ensembles or finite-sized sub-manifolds in
  phase space and not single time-dependent trajectories. Therefore,
  ensemble averages are the natural objects of statistical
  probabilities.  This is the physical origin of coarse-graining which
  is not anymore a mathematical ad hoc assumption.
 The probabilities $P(M)$ of macroscopic measurements
  $\hat{M}$ are given by the ratio $P(M)=W(M)/W$ of these volumes of
  the sub-manifold ${\cal{M}}$ of the microcanonical ensemble with the
  constraint $M$ to the one without.  From this concept all
  equilibrium thermodynamics can be deduced quite naturally {\em
    including the most sophisticated phenomena of phase transitions
    for ``Small'' systems.}
  
  Boltzmann's principle is generalized to non-equilibrium Hamiltonian
  systems with possibly fractal distributions ${\cal{M}}$ in
  $6N$-dim.  phase space by replacing the conventional Riemann
  integral for the volume in phase space by its corresponding
  box-counting volume.  This is equal to the volume of the closure
  $\overline{\cal{M}}$. With this extension the Second Law is derived
  without invoking the thermodynamic limit.
  
  The irreversibility in this approach is due to the replacement of
  the phase-space volume of the fractal sub-manifold ${\cal{M}}$ by
  the volume of its closure $\overline{\cal{M}}$. The physical reason for
  this replacement is that macroscopic measurements cannot distinguish
  ${\cal{M}}$ from $\overline{\cal{M}}$. Whereas the former is not changing
  in time due to Liouville's theorem, the volume of the closure can be
  larger. In contrast to conventional coarse graining the box-counting
  volume is defined in the limit of infinite resolution.  I.e. there
  is no artificial loss of information.
\end{abstract}
\section{Introduction}
Recently the interest in the thermo-statistical behaviour of
non-extensive many-body systems, like atomic nuclei, atomic clusters,
soft-matter, biological systems  ---  and also self-gravitating
astro-physical systems lead to consider thermo-statistics without
using the thermodynamic limit. This is most safely done by going
back to Boltzmann.

Einstein calls Boltzmann's definition of entropy as e.g. written on
his famous epitaph
\begin{equation}
\fbox{\fbox{\vecbm{$S=k$\cdot$lnW$}}}\label{boltzmentr1}\end{equation}
as Boltzmann's principle \cite{einstein05d} from which Boltzmann
was able to deduce thermodynamics. Here $W$ is the number of 
micro-states at given energy $E$ of the $N$-body system in the 
spatial volume $V$:
\begin{eqnarray}
W(E,N,V)&=& tr[\epsilon_0\delta(E-\hat H_N)]\label{partitsum}\\
tr[\delta(E-\hat H_N)]&=&\int_{\{q\subset V\}}{\frac{1}{N!}
\left(\frac{d^3q\;d^3p}
{(2\pi\hbar)^3}\right)^N\delta(E-\hat H_N)},\label{phasespintegr}
\end{eqnarray} 
$\epsilon_0$ is a suitable energy constant to make $W$ dimensionless,
$\hat H_N$ is the $N$-particle Hamilton-function and the $N$
positions $q$ are restricted to the volume $V$, whereas the momenta
$p$ are unresticted.  In what follows, we remain on the level of
classical mechanics. The only reminders of the underlying quantum
mechanics are the measure of the phase space in units of $2\pi\hbar$
and the factor $1/N!$ which respects the indistinguishability of the
particles (Gibbs paradoxon). In contrast to Boltzmann
\cite{boltzmann1877,boltzmann1884} who used the principle only for
dilute gases and to Schr\"odinger \cite{schroedinger44}, who thought
equation (\ref{boltzmentr1}) is useless otherwise, I take the
principle as {\em the fundamental, generic definition of entropy}. In
a recent book \cite{gross174} cf. also \cite{gross173,gross175} I
demonstrated that this definition of thermo-statistics works well
especially also at higher densities and at phase transitions without
invoking the thermodynamic limit.

Before we proceed we must comment on Einstein's attitude to the principle
\cite{pais82}): Originally, Boltzmann called $W$ the
``Wahrscheinlichkeit'' (probability), i.e. the relative time a system
spends (along a time-dependent path) in a given region of $6N$-dim.
phase space. Our interpretation of $W$ to be the number of
``complexions'' (Boltzmann's second interpretation) or quantum states
(trace) with the same energy was criticized by Einstein
\cite{einstein05d} as artificial. It is exactly that criticized
interpretation of $W$ which I use here and which works so excellently
\cite{gross174}.  In section \ref{einsteinEPS} I will come back to this
fundamental point.

After succeeding to deduce  equilibrium statistics including all phenomena 
of phase transitions from Boltzmann's principle even for ``Small'' systems,
i.e. non-extensive many-body systems, it is challenging to explore how
far  this ``most conservative  and restrictive way to thermodynamics'' 
\cite{bricmont00} is able to describe also the {\em approach} of
(eventially ``Small'') systems to equilibrium and the Second Law of
Thermodynamics.

Thermodynamics describes the development of {\em macroscopic}
features of many-body systems without specifying them microscopically
in all details. Before we address the Second Law, we have to clarify what
we mean with the label ``macroscopic observable''.
\section{Measuring a macroscopic observable, the `` EPS-formulation ''}
\label{EPSformulation}
A single point $\{q_i(t),p_i(t)\}_{i=1\cdots N}$ in the $N$-body phase
space corresponds to a detailed specification of the system with all
degrees of freedom (d.o.f) completely fixed at time $t$ (microscopic
determination).  Fixing only the total energy $E$ of an $N$-body
system leaves the other ($6N-1$)-degrees of freedom unspecified.  A
second system with the same energy is most likely not in the same
microscopic state as the first, it will be at another point in phase
space, the other d.o.f. will be different. I.e. the measurement of the
total energy $\hat{H}_N$, or any other macroscopic observable
$\hat{M}$, determines a ($6N-1$)-dimensional {\em sub-manifold}
${\cal{E}}$ or ${\cal{M}}$ in phase space. All points in $N$-body
phase space consistent with the given value of $E$ and volume $V$,
i.e. all points in the ($6N-1$)-dimensional sub-manifold
${\cal{E}}(N,V)$ of phase space are equally consistent with this
measurement. ${\cal{E}}(N,V)$ is the microcanonical ensemble. This
example tells us that {\em any macroscopic measurement is incomplete
and defines a sub-manifold of points in phase space not a single
point}. An additional measurement of another macroscopic quantity
$\hat{B}\{q,p\}$ reduces ${\cal{E}}$ further to the cross-section
${\cal{E}\cap\cal{B}}$, a ($6N-2$)-dimensional subset of points in
${\cal{E}}$ with the volume:
\begin{equation}
W(B,E,N,V)=\frac{1}{N!}\int{\left(\frac{d^3q\;d^3p}
{(2\pi\hbar)^3}\right)^N\epsilon_0\delta(E-\hat H_N\{q,p\})\;
\delta(B-\hat B\{q,p\})}
\label{integrM}\end{equation}
If $\hat H_N\{q,p\}$ as also $\hat B\{q,p\}$ are continuous
differentiable functions of their arguments, what we assume in the
following, ${\cal{E}}\cap{\cal{B}}$ is closed. In the following we use
$W$ for the Riemann or Liouville volume of a many-fold.

Microcanonical thermo{\em statics} gives the probability $P(B,E,N,V)$
to find the $N$-body system in the sub-manifold
${\cal{E}\cap\cal{B}}(E,N,V)$:
\begin{equation}
P(B,E,N,V)=\frac{W(B,E,N,V)}{W(E,N,V)}=e^{\ln[W(B,E,N,V)]-S(E,N,V)}
\label{ EPS }\end{equation}
This is what Krylov seems to have had in mind \cite{krylov79} and what
I will call the ``ensemble probabilistic formulation of statistical
mechanics ($EPS$) ''.

Similarly thermo{\em dynamics} describes the development of some
macroscopic observable $\hat{B}\{q_t,p_t\}$ in time of a system which
was specified at an earlier time $t_0$ by another macroscopic
measurement $\hat{A}\{q_0,p_0\}$.  It is related to the volume of the
sub-manifold
${\cal{M}}(t)={\cal{A}}(t_0)\cap{\cal{B}}(t)\cap{\cal{E}}$:
\begin{equation} W(A,B,E,t)=\frac{1}{N!}\int{\left(\frac{d^3q_t\;d^3p_t}
{(2\pi\hbar)^3}\right)^N\delta(B-\hat B\{q_t,p_t\})\;
\delta(A-\hat A\{q_0,p_0\})\;\epsilon_0\delta(E-\hat H\{q_t,p_t\})},
\label{wab}
\end{equation}
where $\{q_t\{q_0,p_0\},p_t\{q_0,p_0\}\}$ is the set of trajectories
solving the Hamilton-Jacobi equations
\begin{equation}
\dot{q}_i=\frac{\partial\hat H}{\partial p_i},\hspace{1cm}
\dot{p}_i=-\frac{\partial\hat H}{\partial q_i},\hspace{1cm}i=1\cdots N
\end{equation}
with the initial conditions $\{q(t=t_0)=q_0;\;p(t=t_0)=p_0\}$.  For a
very large system with $N\sim 10^{23}$ the probability to find a given
value $B(T)$, $P(B(t))$, is usually sharply peaked as function of $B$.
Ordinary thermodynamics treats systems in the thermodynamic limit
$N\to\infty$ and gives only $<\!\!B(t)\!\!>$.  However, here we are
interested to formulate the Second Law for ``Small'' systems i.e.  we
are interested in the whole distribution $P(B(t))$ not only in its
mean value $<\!\!B(t)\!\!>$.  Thermodynamics does {\em not} describe
the temporal development of a {\em single} system (single point in the
$6N$-dim phase space).

There is an important property of macroscopic measurements: Whereas
the macroscopic constraint $\hat{A}\{q_0,p_0\}$ determines (usually) a
compact region ${\cal{A}}(t_0)$ in \{$q_0,p_0$\} this does not need to
be the case at later times $t\gg t_0$: ${\cal{A}}(t)$ defined by
${\cal{A}}\{q_0\{q_t,p_t\},p_0\{q_t,p_t\}\}$ might become a {\em
  fractal} i.e. ``spaghetti-like'' manifold as a function of
$\{q_t,p_t\}$ in ${\cal{E}}$ at $t\to \infty$ and loose compactness.

This can be expressed in mathematical terms: There exist series of
points $\{a_n\}\in{\cal{A}}(t)$ which converge to a point $a_\infty$
which is {\em not} in ${\cal{A}}(t)$. E.g. such points $a_\infty$ may
have intruded from the phase space complimentary to ${\cal{A}}(t_0)$.
Illustrative examples for this evolution of an initially compact
sub-manifold into a fractal set are the baker transformation discussed
in this context by ref. \cite{fox98,gilbert00}. Then no macroscopic
(incomplete) measurement at time $t$ can resolve $a_\infty$ from its
immediate neighbors $a_n$ in phase space with distances
$|a_n-a_\infty|$ less then any arbitrary small $\delta$. In other
words, {\em at the time $t\gg t_0$ no macroscopic measurement with its
  incomplete information about $\{q_t,p_t\}$ can decide whether
  $\{q_0\{q_t,p_t\},p_0\{q_t,p_t\}\}\in{\cal{A}}(t_0)$ or not.} I.e.
any macroscopic theory like thermodynamics can only deal with the {\em
  closure} of ${\cal{A}}(t)$. If necessary, the sub-manifold
${\cal{A}}(t)$ must be artificially closed to $\overline{\cal{A}}(t)$
as developed further in section \ref{fractSL}.  {\em Clearly, in this
  approach this is the physical origin of irreversibility.} We come
back to this in section \ref{fractSL}.
\section{On Einstein's objections against the EPS-probability} 
\label{einsteinEPS}
According to Abraham Pais: ``Subtle is the Lord''\cite{pais82},
Einstein was critical with regard to the definition of relative
probabilities by eq.\ref{ EPS }, Boltzmann's counting of
``complexions''. He considered it as artificial and not corresponding
to the immediate picture of probability used in the actual problem:
``The word probability is used in a sense that does not conform to its
definition as given in the theory of probability.  In particular,
cases of equal probability are often hypothetically defined in
instances where the theoretical pictures used are sufficiently
definite to give a deduction rather than a hypothetical assertion''
\cite{einstein05d}.  He preferred to define probability by the
relative time a system (a trajectory of a single point moving with
time in the $N$-body phase space) spends in a subset of the phase
space.  However, is this really the immediate picture of probability
used in statistical mechanics?  This definition demands the ergodicity
of the trajectory in phase space.  As we discussed above,
thermodynamics as any other macroscopic theory handles incomplete,
macroscopic informations of the $N$-body system. It handles,
consequently, the temporal evolution of {\em finite sized
  sub-manifolds} - ensembles - not single points in phase space.  The
{\em typical} outcomes of macroscopic measurements are calculated.
Nobody waits in a macroscopic measurement, e.g. of the temperature,
long enough that an atom can cross the whole system.

In this respect, I think the EPS version of statistical mechanics is
closer to the experimental situation than the duration-time of a
single trajectory. Moreover, in an experiment on a small system like a
nucleus, the excited nucleus, which then may fragment statistically
later on, is produced by a multiple {\em repetition} of scattering
events and statistical averages are taken. No ergodic covering of the
whole phase space by a single trajectory in time is demanded. At the
high excitations of the nuclei in the fragmentation region their
life-time would bo too short for that. This is analogous to the
statistics of a falling ball on a Galton's nail-board where also a
single trajectory is not touching all nails but is random. Only after
many repetitions the smooth binomial distribution is established. As I
am discussing here the Second Law in {\em finite} systems, this is the
correct scenario, not the time average over a single ergodic
trajectory.
\section{Fractal distributions in phase space, Second Law}\label{fractSL}
Here we will first describe a simple working-scheme (i.e. a sufficient
method) which allows to deduce mathematically the Second Law. Later,
we will show how this method is necessarily implied by the reduced
information obtainable by macroscopic measurements.

Let us examine the following Gedanken experiment: Suppose the
probability to find our system at points $\{q_t,p_t\}_1^N$ in phase
space is uniformly distributed for times $t<t_0$ over the sub-manifold
${\cal{E}}(N,V_1)$ of the $N$-body phase space at energy $E$ and
spatial volume $V_1$. At time $t>t_0$ we allow the system to spread
over the larger volume $V_2>V_1$ without changing its energy.  If the
system is {\em dynamically mixing}, the majority of trajectories
$\{q_t,p_t\}_1^N$ in phase space starting from points $\{q_0,p_0\}$
with $q_0\subset V_1$ at $t_0$ will now spread over the larger volume
$V_2$.  Of course the Liouvillean measure of the distribution
${\cal{M}}\{q_t,p_t\}$ in phase space at $t>t_0$ will remain the same
($=tr[{\cal{E}}(N,V_1)]$) \cite{goldstein59}. (The label $\{q_0\subset
V_1\}$ of the integral means that the positions $\{q_0\}_1^N$ are
restricted to the volume $V_1$, the momenta $\{p_0\}_1^N$ are
unrestricted.)
\begin{eqnarray}
\left.tr[{\cal{M}}\{q_t\{q_0,p_0\},p_t\{q_0,p_0\}\}]
\right|_{\{q_0\subset V_1\}}
&=&\int_{\{q_0\{q_t,p_t\}\subset V_1\}}{\frac{1}{N!}\left(\frac{d^3q_t\;d^3p_t}
{(2\pi\hbar)^3}\right)^N\epsilon_0\delta(E-\hat H_N\{q_t,p_t\})}\nonumber\\
&=&\int_{\{q_0\subset V_1\}}{\frac{1}{N!}\left(\frac{d^3q_0\;
d^3p_0}{(2\pi\hbar)^3}\right)^N\epsilon_0\delta(E-\hat H_N\{q_0,p_0\})},\\
\mbox{because of: }\frac{\partial\{q_t,p_t\}}{\partial\{q_0,p_0\}}&=&1.
\end{eqnarray}
But as already argued by Gibbs the distribution ${\cal{M}}\{q_t,p_t\}$
will be filamented like ink in water and will approach any point of
${\cal{E}}(N,V_2)$ arbitrarily close.  ${\cal{M}}\{q_t,p_t\}$ becomes
dense in the new, larger ${\cal{E}}(N,V_2)$ for times sufficiently
larger than $t_0$.  The closure $\overline{{\cal{M}}}$ becomes equal
to ${\cal{E}}(N,V_2)$.  This is clearly expressed by Lebowitz
\cite{lebowitz99,lebowitz99a}.

In order to express this fact mathematically, {\em we have to redefine
  Boltzmann's definition of entropy eq.(\ref{boltzmentr1}) and
  introduce the following fractal ``measure'' for integrals like
  (\ref{phasespintegr}) or (\ref{integrM}):}
\begin{equation}
W(E,N,t\gg t_0)=
\frac{1}{N!}\int_{\{q_0\{q_t,p_t\}\subset V_1\}}{\left(\frac{d^3q_t\;d^3p_t}
{(2\pi\hbar)^3}\right)^N\epsilon_0\delta(E-\hat H_N\{q_t,p_t\})}
\end{equation}
With the transformation:
\begin{eqnarray}
\int{\left(d^3q_t\;d^3p_t\right)^N\cdots}&=&
\int{d\sigma_1\cdots d\sigma_{6N}\cdots}\\
d\sigma_{6N}&:=&\frac{1}{||\nabla\hat H||}
\sum_i{\left(\frac{\partial\hat H}{\partial
q_i}dq_i+\frac{\partial\hat H}{\partial p_i}dp_i\right)}=
 \frac{1}{||\nabla\hat H||}dE\\
||\nabla\hat H||&=&\sqrt{\sum_i{\left(\frac{\partial\hat H}{\partial
q_i}\right)^2+\sum_i{\left(\frac{\partial\hat H}{\partial p_i}\right)^2}}}\\
W(E,N,t\gg t_0)&=&\frac{1}{N!(2\pi\hbar)^{3N}}
\int_{\{q_0\{q_t,p_t\}\subset V_1\}}
{d\sigma_1\cdots d\sigma_{6N-1}
\frac{\epsilon_0}{||\nabla\hat H||}},
\end{eqnarray}
we replace ${\cal{M}}$ by its closure $\overline{\cal{M}}$ and {\em
  define} now:
\begin{equation}
W(E,N,t\gg t_0)\to M(E,N,t\gg t_0):=<\!G({\cal{E}}(N,V_2))\!>
*\mbox{vol}_{box}[{\cal{M}}(E,N,t\gg t_0)],\label{boxM1}
\end{equation}
where $<\!G({\cal{E}}(N,V_2))\!>$ is the average of
$\frac{\epsilon_0}{N!(2\pi\hbar)^{3N}||\nabla\hat H||}$ over the
(larger) manifold ${\cal{E}}(N,V_2)$, and
$\mbox{vol}_{box}[{\cal{M}}(E,N,t\gg t_0)]$ is the box-counting volume
of ${\cal{M}}(E,N,t\gg t_0)$ which the same as the volume of
$\overline{\cal{M}}$, see below.

To obtain $\mbox{vol}_{box}[{\cal{M}}(E,N,t\gg t_0)]$ we cover the
$d$-dim. sub-manifold ${\cal{M}}(t)$, here with $d=(6N-1)$, of the
phase space by a grid with spacing $\delta$ and count the number
$N_\delta\propto \delta^{-d}$ of boxes of size $\delta^{6N}$, which
contain points of ${\cal {M}}$.  Then we determine
\begin{eqnarray}
\mbox{vol}_{box}[{\cal{M}}(E,N,t\gg t_0)]&:=&\underbar{$\lim$}_{\delta\to 0}
\delta^d N_\delta[{\cal{M}}(E,N,t\gg t_0)]\label{boxvol}\\
\lefteqn{\mbox{with }\underbar{$\lim *$}=\inf[\lim *]\mbox{ or
symbolically:}} \nonumber\\
M(E,N,t\gg t_0)&=:&
\displaystyle{B_d\hspace{-0.5 cm}\int}_{\{q_0\{q_t,p_t\}\subset V_1\}}
{\frac{1}{N!}\left(\frac{d^3q_t\;
d^3p_t}{(2\pi\hbar)^3}\right)^N\epsilon_0\delta(E-\hat
H_N)}\label{boxM}\\ 
&\to&\frac{1}{N!}\int_{\{q_t\subset V_2\}}
{\left(\frac{d^3q_t\;d^3p_t}
{(2\pi\hbar)^3}\right)^N\epsilon_0\delta(E-\hat H_N\{q_t,p_t\})}
\nonumber\\
&=&W(E,N,V_2) \ge W(E,N,V_1),
\end{eqnarray}
where $\displaystyle{B_d\hspace{-0.5 cm}\int}$ means that this
integral should be evaluated via the box-counting volume
(\ref{boxvol}) here with $d=6N-1$.

This is illustrated by the following figure
\begin{figure}
\begin{minipage}[t]{6cm}
\begin{center}$V_a$\hspace{2cm}$V_b$\end{center}
\vspace*{0.2cm}
\includegraphics*[bb = 0 0 404 404, angle=-0, width=5.7cm,
clip=true]{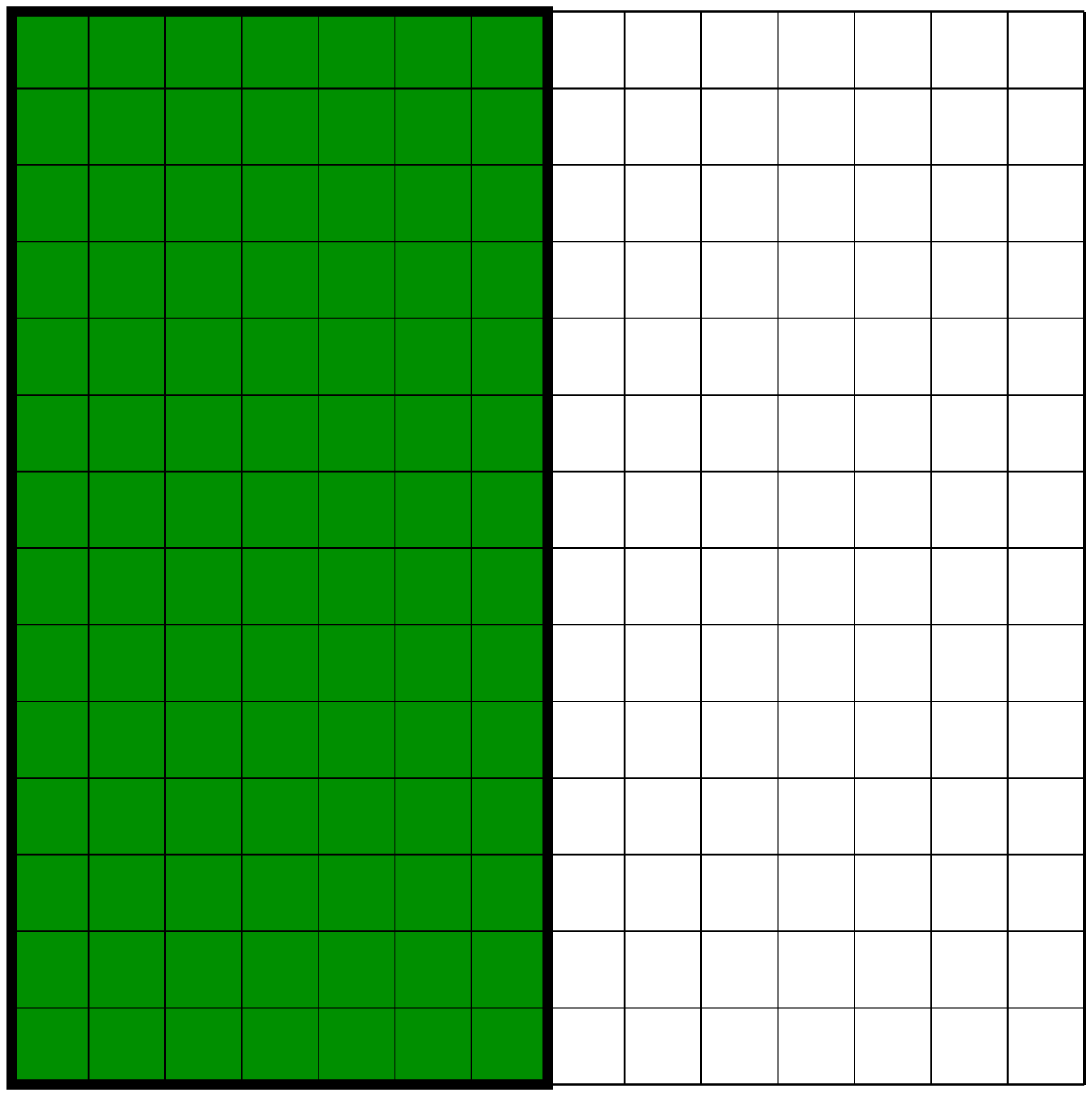}\begin{center}$t<t_0$\end{center}
\end{minipage}\lora\begin{minipage}[t]{6cm}
\begin{center}$V_a+V_b$\end{center}
\includegraphics*[bb = 0 0 428 428, angle=-0, width=6cm,
clip=true]{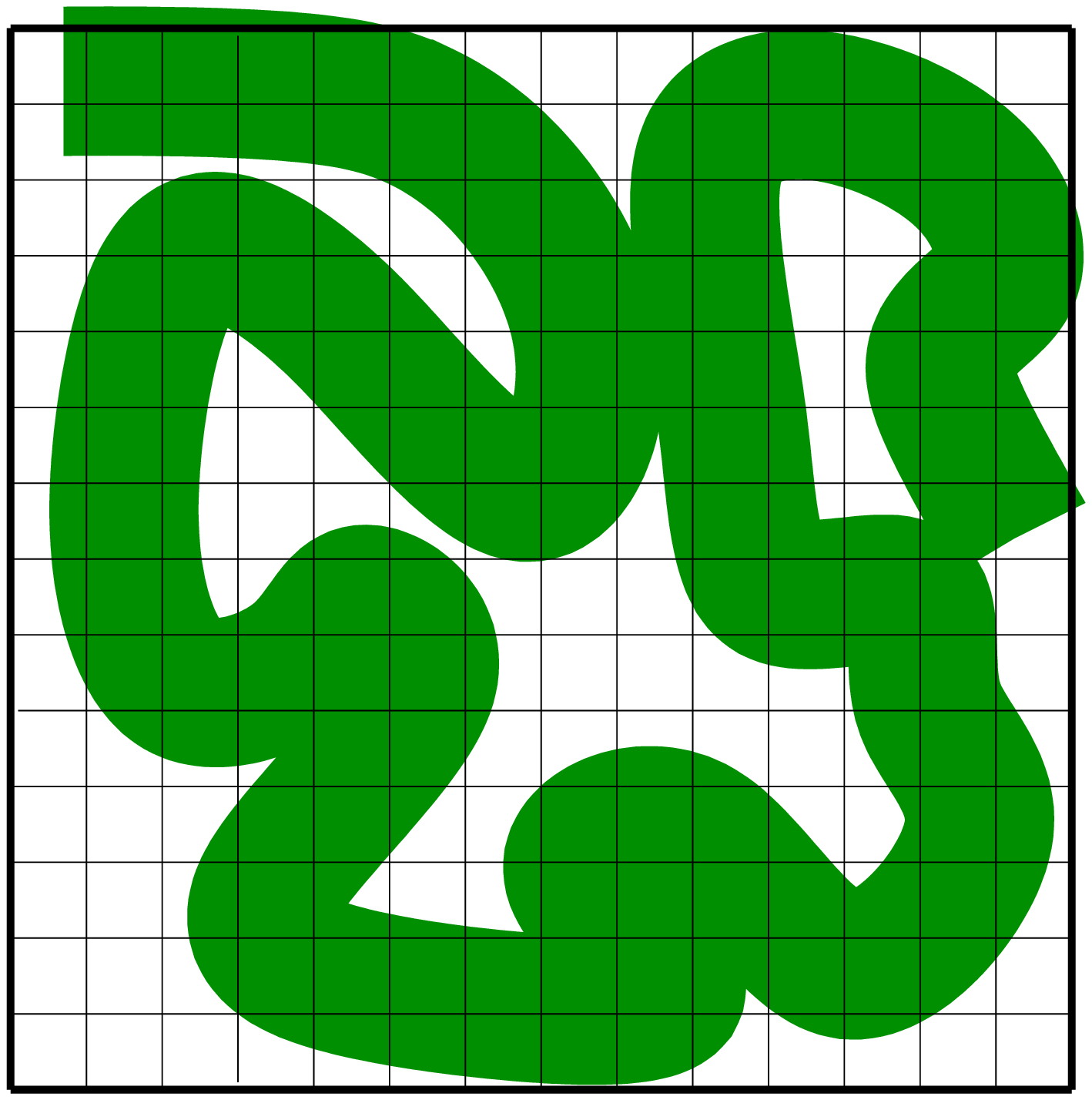}\begin{center}$t>t_0$\end{center}
\end{minipage}
\caption{ The compact set ${\cal{M}}(t_0)$, left side, develops
  into an increasingly folded ``spaghetti''-like distribution in
  phase-space with rising time $t$. This figure shows only the early
  form of the distribution. At much larger times it will become more
  and more fractal and finally dense in the new phase space.  The grid
  illustrates the boxes of the box-counting method.All boxes which
  overlap with ${\cal{M}}(t)$ are counted in $N_\delta$ in
  eq.(\ref{boxvol})}
\end{figure}  
With this extension of eq.(\ref{phasespintegr}) Boltzmann's entropy
(\ref{boltzmentr1}) is at time $t\gg t_0$ equal to the logarithm of
the {\em larger} phase space $W(E,N,V_2)$. {\em This is the Second Law
  of Thermodynamics.}  The box-counting is also used in the definition
of the Kolmogorov entropy, the average rate of entropy gain
\cite{falconer90,crc99}.  Of course still at $t_0$
$\overline{{\cal{M}}(t_0)}={\cal{M}}(t_0)={\cal{E}}(N,V_1)$:
\begin{eqnarray}
M(E,N,t_0)
&=:&\displaystyle{B_d\hspace{-0.5 cm}\int}_{\{q_0\subset V_1\}}
{\frac{1}{N!}\left(\frac{d^3q_0\;
d^3p_0}{(2\pi\hbar)^3}\right)^N\epsilon_0\delta(E-\hat H_N)}\\
&\equiv&\int_{\{q_0\subset V_1\}}
{\frac{1}{N!}\left(\frac{d^3q_0\;
d^3p_0}{(2\pi\hbar)^3}\right)^N\epsilon_0\delta(E-\hat H_N)}\nonumber\\
&=&W(E,N,V_1).
\end{eqnarray} 

The box-counting volume is analogous to the standard method to
determine the fractal dimension of a set of points \cite{falconer90}
by the box-counting dimension:
\begin{equation}
\dim_{box}[{\cal{M}}(E,N,t\gg t_0)]:=\underbar{$\lim$}_{\delta\to 0}
\frac{\ln{N_\delta[{\cal{M}}(E,N,t\gg t_0)]}}{-\ln{\delta}}\\
\end{equation}

Like the box-counting dimension, $\mbox{vol}_{box}$ has the
peculiarity that it is equal to the volume of the smallest {\em
  closed} covering set. E.g.: The box-counting volume of the set of
rational numbers $\{{\bf Q}\}$ between $0$ and $1$, is
$\mbox{vol}_{box}\{{\bf Q}\}=1$, and thus equal to the measure of the
{\em real} numbers , c.f.  Falconer \cite{falconer90} section 3.1.
This is the reason why $\mbox{vol}_{box}$ is not a measure in its
mathematical definition because then we should have
\begin{equation}
\mbox{vol}_{box}\left[\sum_{i\subset\{\bf  Q\}}({\cal{M}}_i)\right]=
\sum_{i\subset\{\bf  Q\}}\mbox{vol}_{box}[{\cal{M}}_i]=0,
\end{equation}
therefore the quotation marks for the box-counting ``measure''.

Coming back to the the end of section (\ref{EPSformulation}), the
volume $W(A,B,\cdots,t)$ of the relevant ensemble, the {\em closure}
$\overline{{\cal{M}}(t)}$ must be ``measured'' by something like the
box-counting ``measure'' (\ref{boxvol},\ref{boxM}) with the
box-counting integral $\displaystyle{ B_d\hspace{-0.5 cm}\int}$, which
must replace the integral in eq.(\ref{phasespintegr}). Due to the fact
that the box-counting volume is equal to the volume of the smallest
closed covering set, the new, extended, definition of the phase-space
integral eq.(\ref{boxM}) is for compact sets like the equilibrium
distribution ${\cal{E}}$ identical to the old one
eq.(\ref{phasespintegr}) and nothing changes for equilibrium
statistics. Therefore, one can simply replace the old
Boltzmann-definition of the number of complexions and with it of the
entropy by the new one (\ref{boxM}).
\section{Conclusion}

Macroscopic measurements $\hat{M}$ determine only a very few of all
$6N$ d.o.f.  Any macroscopic theory like thermodynamics deals with the
{\em volumes} $M$ of the corresponding closed sub-manifolds
$\overline{\cal{M}}$ in the $6N$-dim. phase space not with single
points.  The averaging over ensembles or finite sub-manifolds in phase
space becomes especially important for the microcanonical ensemble of
a {\em finite} system.

Because of this necessarily coarsed information, macroscopic
measurements, and with it also macroscopic theories are unable to
distinguish fractal sets ${\cal{M}}$ from their closures
$\overline{\cal{M}}$. Therefore, I make the conjecture: the proper
manifolds determined by a macroscopic theory like thermodynamics are
the closed $\overline{\cal{M}}$. However, an initially closed subset
of points at time $t_0$ does not necessarily evolve again into a
closed subset at $t\gg t_0$. I.e. the closure operation and the
$t\to\infty$ limit do not commute, and the macroscopic dynamics
becomes irreversible.

Here is the origin of the misunderstanding by the famous reversibility
paradoxes which were invented by Loschmidt \cite{loschmidt76} and
Zermelo \cite{zermelo96,zermelo97} and which bothered Boltzmann so
much \cite{cohen97,cohen00}. These paradoxes address to trajectories
of {\em single points} in the $N$-body phase space which must return
after Poincarre's recurrence time or which must run backwards if all
momenta are exactly reversed. Therefore, Loschmidt and Zermelo
concluded that the entropy should decrease as well as it was
increasing before. The specification of a single point demands of
course a {\em microscopic exact} specification of all $6N$ degrees of
freedom not a determination of a few macroscopic degrees of freedom
only. No entropy is defined for a single point.

This way various non-trivial limiting processes can be avoided.
Neither does one invoke the thermodynamic limit of a homogeneous
system with infinitely many particles nor does one rely on the ergodic
hypothesis of the equivalence of (very long) time averages and
ensemble averages.  {\em The use of ensemble averages is justified
  directly by the very nature of macroscopic (incomplete)
  measurements}. Coarse-graining appears as natural consequence of
this. The box-counting method mirrors the averaging over the
overwhelming number of non-determined degrees of freedom. Of course, a
fully consistent theory must use this averaging explicitly. Then one
would not depend on the order of the limits $\lim_{\delta\to
  0}\lim_{t\to\infty}$ as it was tacitly assumed here. Presumably, the
rise of the entropy can then be already seen at finite times when the
fractality of the distribution in phase space is not yet fully
developed. The coarse-graining is no more any mathematical ad hoc
assumption.  Moreover the Second Law is in the EPS-formulation of
statistical mechanics not linked to the thermodynamic limit as was
thought up to now \cite{lebowitz99,lebowitz99a}.
\section{Acknowledgment}
Thanks to A.Ecker for mathematical advises.
I am especially grateful to E.G.D.Cohen for his interest and his 
constructive criticism of the ideas developed here. He turned my 
attention to the fundamental discussions in Pais' book about
Einsteins view of statistical mechanics. It was through the many 
letters exchanged with Cohen  that I realized how much my way 
to statistical mechanics was not so self-evident as I thought and 
departs from the main stream founded by Boltzmann and Einstein. 
He convinced me to express the fundamental ideas in detail. 
I hope this paper serves this purpose. The many applications to
realistic problems especially to phase transitions in ``Small'' 
systems are given in my book \cite{gross174}. 

%%%\bibliographystyle{unsrt}%{alpha}%{plain} %{unsrt}
%%%\bibliography{gross,othbiba,othbibb,othbibcd,othbibe,othbibf,othbibg,othbibh,othbibij,othbibk,othbibl,othbibm,othbibn,othbibo,othbibp,othbibr,othbibs,othbibt,othbibuw,othbibxz}
%%%\end{document}

\end{document}